\documentclass[aps, prd, showpacs, showkeys, preprint, eqsecnum, floatfix, superscriptaddress, nofootinbib, longbibliography]{revtex4-2}

\usepackage{hyperref}
\usepackage{amssymb}
\usepackage{amsmath}
\usepackage{amsthm}
\usepackage{bm}
\usepackage{calrsfs}
\usepackage{color}
\usepackage{graphicx}
\usepackage[normalem]{ulem} 
\usepackage{relsize}
\usepackage{multirow}

\newcommand{\bq}{\begin{eqnarray}}
\newcommand{\eq}{\end{eqnarray}}
\newcommand{\bqn}{\begin{eqnarray*}}
\newcommand{\eqn}{\end{eqnarray*}}

\newcommand{\ZZ}{{\rm Z}\hskip-.3em{\rm Z}}

\newcommand{\calo}{{\cal O}}

\begin{document}
\title{Monte Carlo evaluation of the continuum limit of the two-point function of two  
Euclidean Higgs real scalar fields subject to affine quantization} 

\author{Riccardo Fantoni}
\email{riccardo.fantoni@posta.istruzione.it}
\affiliation{Universit\`a di Trieste, Dipartimento di Fisica, strada
  Costiera 11, 34151 Grignano (Trieste), Italy}


\author{John R. Klauder}
\email{klauder@ufl.edu}
\affiliation{Department of Physics and Department of Mathematics \\
University of Florida,   
Gainesville, FL 32611-8440}

\date{\today}

\begin{abstract}
We study canonical and affine versions of the quantized covariant Euclidean Higgs 
scalar field-theory for two real fields on four dimensional lattices through the Monte 
Carlo method. We calculate the two-point function near the continuum limit at finite 
volume.
\end{abstract}



\maketitle
\section{Introduction}
\label{sec:introduction}

The aim of this work is to find out what affine quantization 
\cite{Klauder2020c,Klauder2000} does to a classical field-theory for two real scalar 
fields, or equivalently a complex scalar field, of mass $m$ subject to the Mexican-hat, 
Higgs potential, rather than canonical quantization \citep{Dirac}. To this aim we will 
compare the two-point function of the two fields in the two frameworks. 

In particular in this paper we try to understand in what ways an affine quantization
is similar as well as dissimilar from a canonical quantization. We add
that some non-free real scalar fields have already been observed and that canonical
quantization fails for several non-renormalizable fields, such as $(\phi^{12})_3$ 
\cite{Fantoni2020} and $(\phi^4)_4$ \cite{Fantoni2020a}. The key to that result
is the introduction of a highly unusual, additional, non-quadratic, term
that is dictated by affine quantization. While affine quantization employs
an additional term, that particular term formally disappears when the Planck constant 
$\hbar\to 0$, which makes it a plausible modification of the quadratic terms of 
traditional free real scalar fields in order to extend acceptable quantization of 
traditional non–renormalizable models. 
\cite{Frohlich1982,Weisz1987,Weisz1988,Weisz1989}

This work should be considered as a follow up of our previous work \cite{Fantoni2020b} 
where the two-point function of a single Euclidean free real scalar field subject to 
affine quantization was found through Monte Carlo (MC) methods. In particular in that 
work we found that the vacuum expectation value of the field diverges in the continuum 
limit. This shortcoming is expected to disappear in the present case of a complex field 
$\varphi=\phi_1+i\phi_2$. In fact, in this case, one can go ``slowly'' ``around''  the  
peak at $\varphi=0$ with no need of ``jumps'' \cite{Fantoni2020b}.

The covariant Euclidean action in canonical quantization \citep{Dirac} is
\footnote{Note however that Eq. (\ref{eq:V}) can be simplified to 
$V=g\left[\left(\phi_1^2+\phi_2^2\right)-A^2\right]^2+\mbox{constant}$, where $A$ 
involves a combination of $\Phi$ and $m$: $A^2=\Phi^2-m^2/4g$.}
\bq \label{eq:c-action}
S^{(c)}[\phi_1,\phi_2]&=&\mathop{\mathlarger{\mathlarger{\int}}}\left\{\frac{1}{2}
\sum_{\mu=0}^s\left[\left(\frac{\partial\phi_1(x)}{\partial x_\mu}\right)^2+
\left(\frac{\partial\phi_2(x)}{\partial x_\mu}\right)^2\right]+
V(\phi_1(x),\phi_2(x))\right\}\,d^nx,\\ \label{eq:V}
V(\phi_1,\phi_2)&=&\frac{1}{2}m^2\left(\phi_1^2+\phi_2^2\right)+
g\left[\left(\phi_1^2+\phi_2^2\right)-\Phi^2\right]^2,
\eq
with $x=(x_0,x_1,\ldots,x_s)=(x_0,\vec{x})$ for $s$ spatial dimensions and $n=s+1$ for 
the number of space-time dimensions with $x_0=ct$, where $c$ is the speed of light 
constant and $t$ extends from zero to $\hbar\beta$ with $\beta=1/k_BT$, $k_B$ being the 
Boltzmann constant and $T$ the absolute temperature. We will work at $s=3$. And $V$ is 
the self-interaction potential density corresponding to an interacting Higgs theory 
with a bare mass $m$ and a bare coupling $g$.

The covariant Euclidean action in affine quantization \cite{Klauder2020c,Klauder2000} 
is
\bq \nonumber
S^{(a)}[\phi_1,\phi_2]=\mathop{\mathlarger{\mathlarger{\int}}}
\left\{\vphantom{\frac{1}{2}}\right.&&\frac{1}{2}
\sum_{\mu=0}^s\left[\left(\frac{\partial\phi_1(x)}{\partial x_\mu}\right)^2+
\left(\frac{\partial\phi_2(x)}{\partial x_\mu}\right)^2\right]+\\ \label{eq:a-action}
&&\left.\frac{3}{8}\frac{\delta^{2s}(0)\hbar^2}{\phi_1^2(x)+\phi_2^2(x)+\epsilon}
+V(\phi_1(x),\phi_2(x))\right\}\,d^nx,
\eq
where $\epsilon>0$ is a parameter used to regularize the ``$3/8$'' extra term stemming 
from considering the complex field $\varphi(x)=\phi_1(x)+i\phi_2(x)$ and the momentum 
field $\pi(x)=-i\hbar\partial/\partial\varphi(x)$ as the two conjugate canonical 
variables (see Appendix A in \cite{Fantoni2020}) and $\delta$ is a Dirac delta 
function. In this case the Hamiltonian density formally contains a divergent term,
\footnote{The divergent integral $\int_{-N}^N\,d\phi/\phi^2$, can be made finite simply 
by a regularized integral such as
$\int{\sum'}_{n=-N}^N(1/3r^2)[(n+1)^2\cos^2(k)+n^2\sin^2(k)\cos^2(k’)+(n-1)^2\sin^2(k)\sin^2(k’)]^{-1}\,dr^3\,dk\,dk'$ where the prime over the sum indicates that we are 
considering a periodic closure $-N,N$ for the three terms in square brackets.} 
in the total potential density 
${\cal V}(\phi)=\frac{3}{8}\delta^{2s}(0)\hbar^2/(\phi_1^2+\phi^2_2+\epsilon)+V(\phi)$, 
in the continuum, but the field theory can be regularized and treated on a lattice, and 
the {\sl approach} toward the continuum will be taken under exam in this work. In the 
following we will use natural units with $c=\hbar=k_B=1$. 

In our previous works we studied the single real scalar field non-renormalizable 
canonical cases with $V(\phi)=\frac{1}{2}m^2\phi^2+g\phi^4$ \cite{Fantoni2020a} in 
$s=3$ and $\frac{1}{2}m^2\phi^2+g \phi^{12}$ in $s=2$ \cite{Fantoni2020}, where $g$ is 
the bare coupling constant. And we showed that the corresponding affine cases are 
indeed renormalizable.

MC \cite{Kalos-Whitlock,Metropolis} is the numerical method of choice to treat 
multidimensional integrals of high dimensions and, therefore, is especially useful to 
compute path integrals. We will use it to study the two-point function of the Euclidean 
action of two real scalar field in affine quantization. Our estimate of the path 
integrals will be generally subject to three sources of numerical uncertainties: The 
one due to the statistical errors, the one due to the space-time discretization, and 
the one due to the finite-size effects. Of these, the statistical errors scale like 
$M^{-1/2}$ where $M$ is the computer time, the discretization of space-time is 
responsible for the distance from the {\sl continuum limit} (which corresponds to a 
lattice spacing $a\to 0$), and the finite-size effects stems from the necessity to 
approximate the infinite space system with one in a periodic box of volume $L^s$ 
with $L=Na$ being the box side, subject to $N$ discretization points. The finite-size 
effects are due to the distance from the {\sl thermodynamic limit} (which corresponds 
to $N\to\infty$). \cite{Wolff2014}

The work is organized as follows: In section \ref{sec:model} we derive the lattice 
formulation of the field theory needed in the treatment on the computer; in section 
\ref{sec:observables} we describe our computer experiment and introduce the observables 
that will be measured during our simulations; in section \ref{sec:results} we present 
our partial results obtained by working with the two scalar fields $\phi_1$ and 
$\phi_2$ where we encounter ergodicity problems for the affine case; in section 
\ref{sec:exponential} we are able to overcome the ergodicity breakdown observed in the 
previous section and we present our final results for the affine case obtained by 
working with the two scalar fields $\rho=\sqrt{\phi_1^2+\phi_2^2}$ and 
$\theta=\arctan(\phi_2/\phi_1)$ such that $d\phi_1d\phi_2=\rho d\theta d\rho$. Section 
\ref{sec:conclusions} is for final remarks.

\section{The lattice formulation of the field-theory model}
\label{sec:model}

We used a lattice formulation of the field theory. The theory considers a complex 
scalar field $\varphi=\phi_1+i\phi_2$ taking the value $\varphi(x)$ on each site of a 
periodic, hypercubic, $n$-dimensional lattice of lattice spacing $a$ and periodicity 
$L=Na$. The canonical covariant action for the field, Eq. (\ref{eq:c-action}), is then 
approximated by
\bq \nonumber
\frac{S^{(c)}[\phi_1,\phi_2]}{a^n}\approx\frac{1}{2a^2}&\sum_{x,\mu}&\left\{\left[\phi_1(x)-\phi_1(x+e_\mu)\right]^2+\left[\phi_2(x)-\phi_2(x+e_\mu)\right]^2\right\}+\\ \label{eq:pa}
&\sum_x&V(\phi_1(x),\phi_2(x)),
\eq
where $e_\mu$ is a vector of length $a$ in the $+\mu$ direction and we are at a 
temperature $T=1/Na$, in units where Boltzmann constant $k_B=1$. 

Note that in our model the continuous symmetry $\varphi\to e^{i\alpha}\varphi$ breaks 
down spontaneously and the mass spectrum contains a Goldstone boson. The accepted 
signal of a system being in the symmetry broken phase in a finite volume, in the 
absence of a small symmetry breaking term, is not a non-zero order parameter, but 
rather the fact that a product of order parameters, at points $x,y$, tends to a 
non-zero limit with increasing $|x-y|$. To understand the properties of the system at 
finite volume, it is convenient to add a small symmetry breaking term and to work with 
the potential 
\bq
V =g(\phi_1^2+\phi_2^2 - A^2)^2 + (\varepsilon^2/2) \phi_2^2 + \mbox{constant},
\eq
The term proportional to $\varepsilon^2$ ensures that the classical action has a proper 
minimum at the point $\phi_1 = A, \phi_2 = 0$.
The expansion of the potential in powers of $\psi = \phi_1-A$, and $\phi_2$ starts 
with 
\bq
V &=&(M^2/2) \psi^2+ (\varepsilon^2/2) \phi_2^2 +\ldots,\\
M &=& A \sqrt{8 g}.
\eq
The first term represents a free particle of mass $M$, the second a free particle of 
mass $\varepsilon$. The situation is the same as in the case of the free real scalar 
field: the perturbative expansion of the two-point function starts with
\bq
\langle\phi_1(x)\phi_1(y)\rangle &=& A^2+ D(x-y,M,L),\\
\langle\phi_2(x)\phi_2(y)\rangle &=& D(x-y,\varepsilon,L),\\ 
\langle\phi_1(x)\phi_2(y)\rangle &=& 0,
\eq
where $\langle\ldots\rangle$ is the vacuum expectation value (defined in Eq. 
(\ref{eq:expectation})) and $D(z,m,L)$ is the propagator of a free particle of mass $m$ 
on a hypercubic Euclidean box of size $L^n$. For $\varepsilon = 0$, the term 
$D(z,\varepsilon,L)$ reduces to a sum of free massless propagators:
\bq
D(z,0,L)=(1/4\pi^2) \sum_{n_0,n_1,n_2,n_3} 1/[(z_0+n_0 L)^2+\ldots+(z_3+n_3 L)^2],
\eq
where $z=(z_0,z_1,\ldots,z_s)$ and $n_\mu\in\ZZ$ for $\mu=0,1,\ldots,s$, but this 
expression does not make sense because the sum diverges. As long as $\varepsilon$ is 
different from zero, the limit $L\to\infty$ ensures that a single term in the sum 
survives, the one with $n_0 = \ldots = n_3 = 0$, which describes the contribution from 
the Goldstone boson.

Expression (\ref{eq:pa}) needs to be modified for the affine action of Eq. 
(\ref{eq:a-action}). In this case the Dirac delta function is replaced by 
$\delta^{2s}(0)\to a^{-2s}$. Moreover it is convenient the following scaling: 
$\phi_i=a^{-s/2}\bar{\phi}_i$, $\Phi_i=a^{-s/2}\bar{\Phi}_i$, $g=a^s\bar{g}$, and 
$\epsilon=a^{-s}\bar{\epsilon}$ which gives the following discretized approximation for 
the affine action
\bq \nonumber
\frac{S^{(a)}[\bar{\phi}_1,\bar{\phi}_2]}{a^{-s}a^n}\approx\frac{1}{2a^2}&\sum_{x,\mu}&\left\{\left[\bar{\phi}_1(x)-\bar{\phi}_1(x+e_\mu)\right]^2+\left[\bar{\phi}_2(x)-\bar{\phi}_2(x+e_\mu)\right]^2\right\}+\\ \nonumber
&\sum_x&\frac{3}{8}\frac{1}{\bar{\phi}_1^2(x)+\bar{\phi}_2^2(x)+\bar{\epsilon}}+\\ \label{eq:pb}
&\sum_x&\left\{\frac{1}{2}m^2\left(\bar{\phi}_1^2(x)+\bar{\phi}_2^2(x)\right)+\bar{g}\left[\left(\bar{\phi}_1^2(x)+\bar{\phi}_2^2(x)\right)-\bar{\Phi}^2\right]^2\right\}.
\eq

Note that if $g$ is taken different from zero, the relation $g = a^s\bar{g}$          
shows that $\bar{g}$ carries a dimension. Setting $\bar{g} = M^s$, $M$ is of dimension 
mass (we are using natural units $c=\hbar=1$). If $M$ as well as $m$ are kept fixed 
when the cutoff is removed, the model 
contains the two dimension-full parameters $m$ and $M$. The lattice spacing $a$ must be 
small compared to $1/m$ as well as compared to $1/M$ and the box must be large compared 
to $1/M$. Since $\bar{\phi}$ is of dimension $\mbox{mass}^{-1/2}$, the two-point 
function of $\bar{\phi}$ is of the form
\bq
\langle\bar{\phi}_i(x) \bar{\phi}_j(y)\rangle = f_{ij}\{M (x-y), m/M, a M, L M\}/M.
\eq
To approach the continuum limit, the last two argument must be in the range:
$a M \ll 1$,   $L M \ll 1$. The only relevant parameter, apart from the number of 
lattice points, used to regularize the system should be the ratio $m/M$. 
 
We will use the so called ``primitive approximation'' for the action (see Eqs. 
(\ref{eq:pa}) or (\ref{eq:pb})) even if it can be improved in several ways 
\citep{Ceperley1995} in order to reduce the error due to the 
space-time discretization. In reaching to the expression (\ref{eq:pa}) or (\ref{eq:pb}) 
we neglected the term $\propto a^{2n}$ due to the commutator of the kinetic and 
potential parts of the Hamiltonian, in the Baker–Campbell–Hausdorff formula. In 
reaching to the path integral expression this is justified by the Trotter formula.

The vacuum expectation of a functional observable $\calo[\phi_1,\phi_2]$ is
\bq \label{eq:expectation}
\langle\calo\rangle\approx\frac{\int\calo[\phi_1,\phi_2]\exp(-S[\phi_1,\phi_2])\,\prod_{x}d\phi_1(x)\phi_2(x)}{\int\exp(-S[\phi_1,\phi_2])\,\prod_{x}d\phi_1(x)\phi_2(x)}, 
\eq
for a given action $S$.

We will approach the continuum limit by choosing a fixed $L$ and increasing the 
number of discretizations $N$ of each component of the space-time. So that the 
lattice spacing $a=L/N\to 0$. To make contact with the continuum limit, two conditions 
must be met $a \ll 1/m \ll L$ where $1/m$ is the Compton wavelength.

\section{Simulation details and Relevant observables} 
\label{sec:observables}

We want to determine the two-point function
\bq \label{eq:Fxy}
K_{ij}(x,y)=\langle\phi_i(x)\phi_j(y)\rangle,
\eq
where in the affine case we need to replace the fields $\phi_i$ by the scaled fields 
$\bar{\phi}_i$. Replacing $x$ by $x+k$ with $k=a w_n$ with $w_n=(n_0,n_1,\ldots,n_s)$ 
and $n_\mu\in\ZZ$ amounts to a mere relabeling of the lattice points. Hence, due to 
translational invariance, $K(x,y)$ can only depend on the difference between the 
coordinates of the two points and we can define,
\bq \label{eq:tp}
D_{ij}(z)=\frac{1}{L^n}\sum_x K_{ij}(x,x+z)a^n.
\eq
Moreover due to the symmetry $1\leftrightarrow 2$ we will have 
$D_{11}=D_{22}\equiv D_{\rm like}$ and $D_{12}=D_{21}\equiv D_{\rm unlike}$. In our 
simulations we work in periodic space-time (at a temperature $T=1/Na$) so that 
$\phi_i(x_\mu+N)=\phi_i(x_\mu)$ for any $x$, $\mu=0,1,\ldots,s$, and $i=1,2$.

Our MC simulations use the Metropolis algorithm \citep{Kalos-Whitlock,Metropolis} 
to calculate the ensemble average of Eq. (\ref{eq:expectation}) which is a $2N^{n}$ 
multidimensional integral. The simulation is started from the initial condition 
$\phi_i=0$ for $i=1,2$. One MC step consisted in a random displacement of each one of 
the $2N^{n}$ variables $\phi_i(x)$ for $i=1,2$, as follows
\bq \label{eq:move}
\phi_i\rightarrow\phi_i+(2\eta-1)\delta,
\eq
where $\eta$ is a uniform pseudo random number in $[0,1]$ and $\delta$ is the 
amplitude of the displacement. The fields $\phi_i\in(-\infty,\infty)$ for $i=1,2$ and 
$x_\mu\in[0,L]$ for $\mu=0,1,\ldots,s$. Each one of these $2N^{n}$ moves is accepted if 
$\exp(-\Delta S)>\eta$ where $\Delta S$ is the change in the action due to the move  
(it can be efficiently calculated considering how the kinetic part and the 
potential part change by the displacement of a single $\phi_i(x)$)
and rejected otherwise. The amplitude $\delta$ is chosen in such a way to have 
acceptance ratios as close as possible to $1/2$ and is kept constant during the 
evolution of the simulation. One simulation consisted of $M$ MC steps each of which 
consisted in a sweep of $2N^{n}$ displacement moves of all the fields variables. The 
statistical error on the average $\langle\calo\rangle$ will then 
depend on the correlation time necessary to decorrelate the property $\calo$, 
$\tau_\calo$, and will be determined as $\sqrt{\tau_\calo\sigma_\calo^2/(M2N^{n})}$, 
where $\sigma_\calo^2$ is the intrinsic variance for $\calo$.  

\section{Simulation results} 
\label{sec:results}

We worked in units where $c=\hbar=k_B=1$. We chose the regularization parameter of 
the affine quantization term to be $\epsilon=10^{-10}$. 
\footnote{Note that we could as well choose a regularization putting hard walls at 
$\phi_i=\pm\varepsilon$ therefore rejecting MC moves whenever 
$\phi_i\in[-\varepsilon,\varepsilon]$, for $i=1,2$.}

\begin{figure}[htbp]
\begin{center}
\includegraphics[width=12cm]{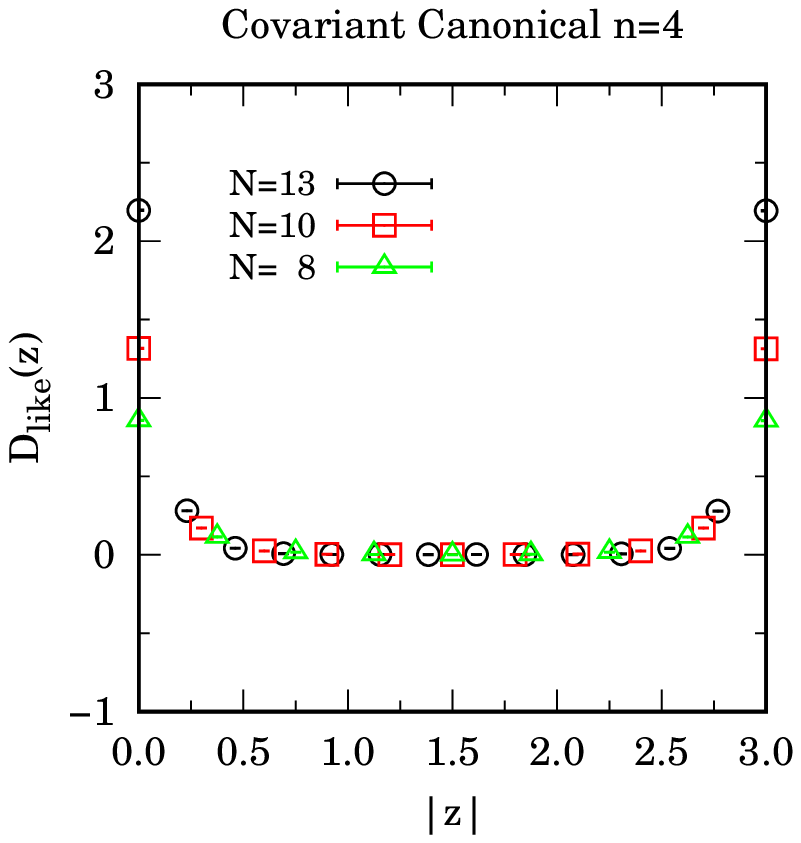}\\
\includegraphics[width=12cm]{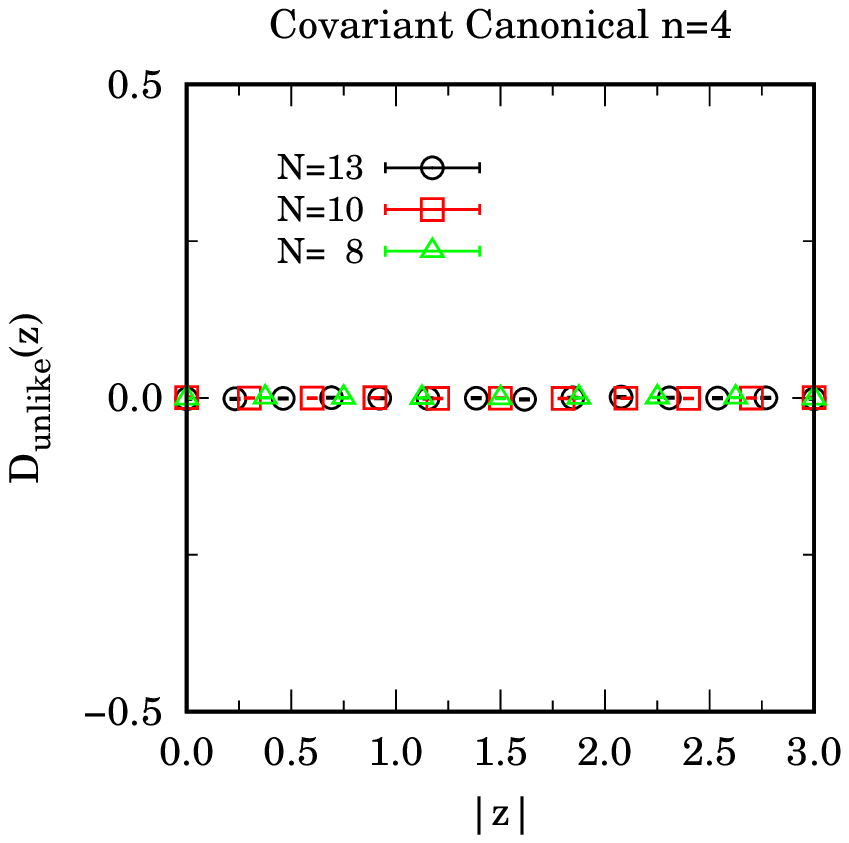}
\end{center}
\caption{(color online) Two-point functions, $D_{\rm like}(z)$ (top panel) and 
$D_{\rm unlike}(z)$ (bottom panel), of Eq. (\ref{eq:tp}) for the complex 
scalar Higgs field $\varphi=\phi_1+i\phi_2$ subject to canonical quantization with a 
self-interaction potential density of the form 
$V(\phi)=\frac{1}{2}m^2(\phi_1^2+\phi_2^2)+g(\phi_1^2+\phi_2^2-\Phi^2)^2$ in Eq. 
(\ref{eq:a-action}) with $m=1, g=1, \Phi=1, L=3$ ($\Pi_c^2=3/4$) and increasing 
$N=8,10,13$. On the abscissa axis we have $|z|=\sqrt{z_0^2+z_1^2+\ldots+z_s^2}$ which 
is a length.}
\label{fig:tp-c}
\end{figure}

In Fig. \ref{fig:tp-c} we show $D_{\rm like}(z)$ and $D_{\rm unlike}(z)$ as obtained 
for  $m=1, g=1, \Phi=1, L=3$ and three choices of $N$, in the canonical scenario. One 
can then see the approach to the continuum of the two-point functions of the canonical 
model. From the figure we can see that the unlike two-point function is zero over the 
whole space-time volume. This can be explained observing that during the random-walk 
the field will be localized around the minima of the potential density so that 
$\phi_1^2+\phi_2^2\approx\Pi_c^2$ with $\Pi_c$ the radius of the minima ring, the 
circle of vacua, around the origin $\varphi=0$, which is a function of $m$, $g$ and 
$\Phi$:
\bq
\Pi_c^2=\frac{4g\Phi^2-m^2}{4g}.
\eq
So that the Higgs potential density in the action does not actually
contribute to correlate the two fields $\phi_i$ for $i=1,2$. Moreover, the expectation 
values $\langle\phi_i\rangle=0$ for $i=1,2$ because the complex field $\varphi$ tends 
to rotate around the origin on the minima ring. The approach to the continuum is 
manifested through increasing values of $D_{\rm like}(0)$ with increasing $N$.
For our choice of the parameters $m^2<4g\Phi^2$ and we must have symmetry breaking
\cite{Weisz1987,Weisz1988,Weisz1989}, with the circle of vacua having a radius 
different from zero. The renormalized coupling constant \cite{Fantoni2020a} was found 
to be: 
$g_R=-0.0069(6)$ for $N=8$,
$g_R=-0.0006(4)$ for $N=10$,
$g_R=+0.0000(5)$ for $N=13$. Since $g_R$ must be non-negative, by Lebowitz inequality, 
our results signal a free trivial system in the continuum limit. 

\begin{figure}[htbp]
\begin{center}
\includegraphics[width=12cm]{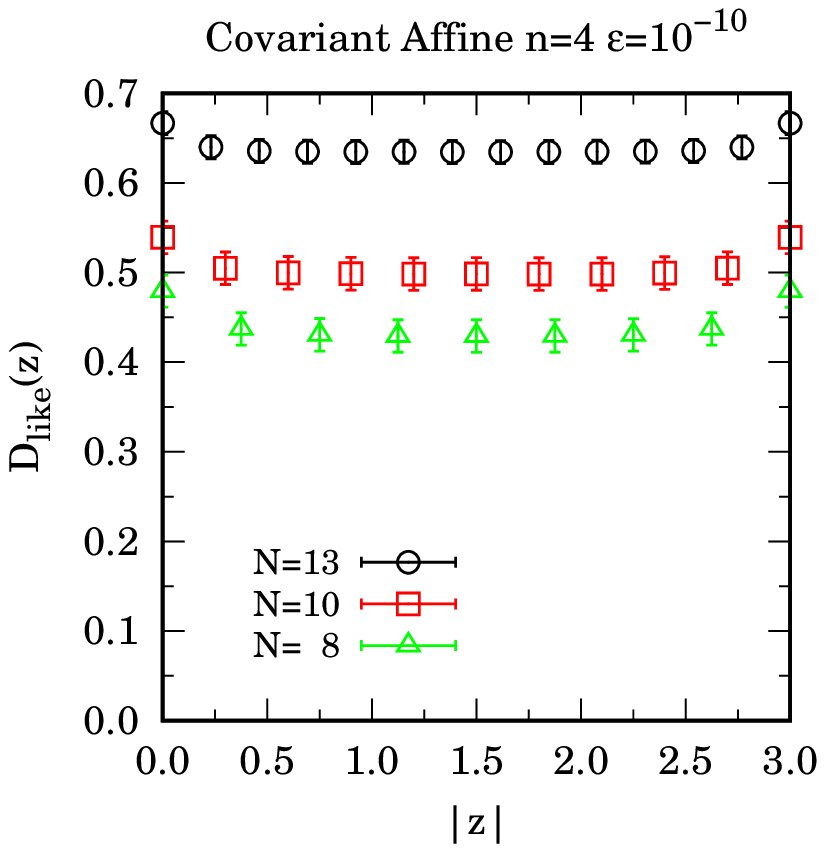}\\
\includegraphics[width=12cm]{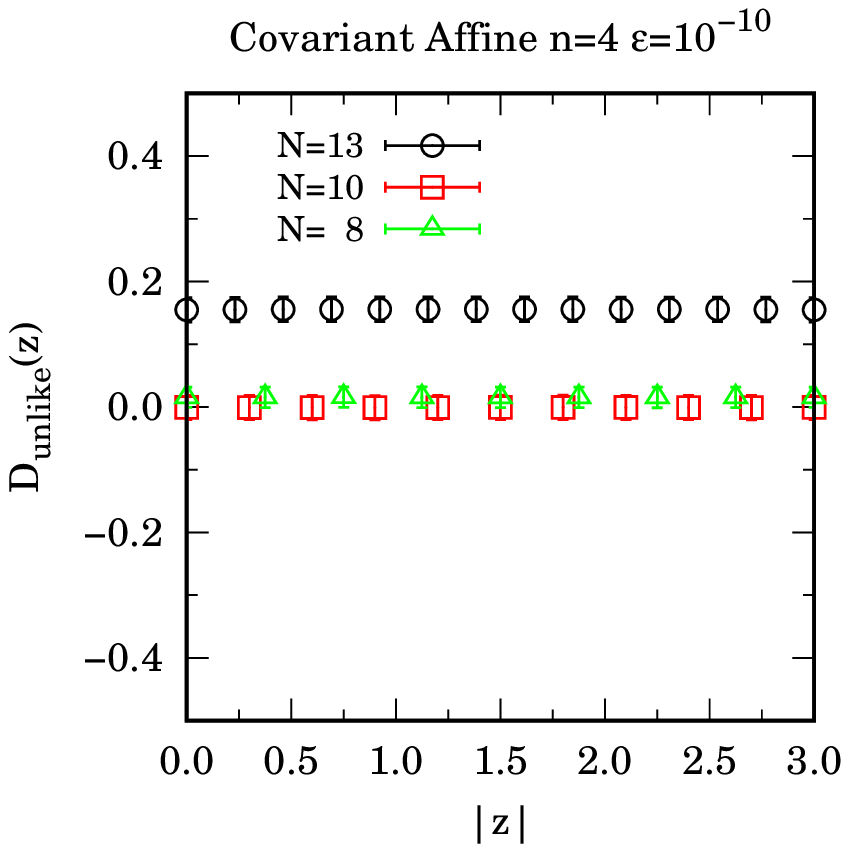}
\end{center}
\caption{(color online) Two-point functions, $D_{\rm like}(z)$ (top panel) and 
$D_{\rm unlike}(z)$ (bottom panel), of Eq. (\ref{eq:tp}) for the complex 
scalar Higgs field $\bar{\varphi}=\bar{\phi}_1+i\bar{\phi}_2$ subject to affine 
quantization with a self-interaction potential density of the form 
$V=\frac{1}{2}m^2(\bar{\phi}_1^2+\bar{\phi}_2^2)+g(\bar{\phi}_1^2+\bar{\phi}_2^2-\Phi^2)^2$ in Eq. 
(\ref{eq:a-action}) with $m=1, \bar{g}=1, \bar{\Phi}=1, L=3, \epsilon=10^{-10}$ 
($\bar{\Pi}_a^2\approx 0.955410$) in Eq. (\ref{eq:pb}) and increasing $N=8,10,13$. The 
simulation used $M=10^7$ MC steps. On the abscissa axis we have 
$|z|=\sqrt{z_0^2+z_1^2+\ldots+z_s^2}$ which is a length.}
\label{fig:tp-a}
\end{figure}

\begin{figure}[htbp]
\begin{center}
\includegraphics[width=12cm]{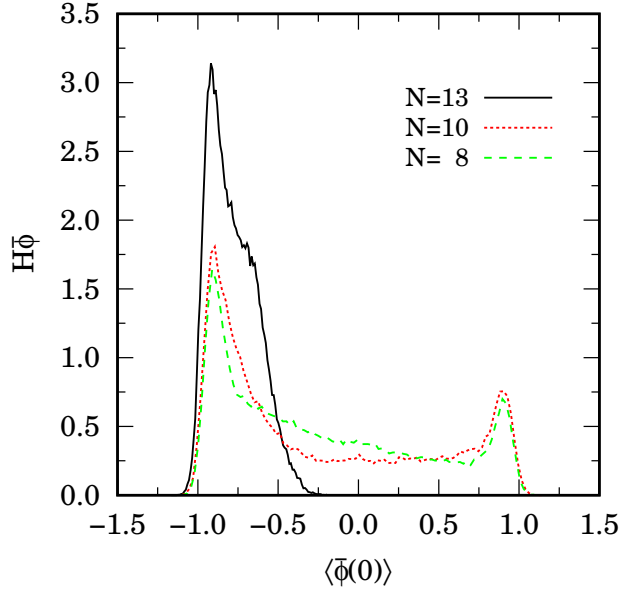}
\end{center}
\caption{(color online) Histogram of $\langle\bar{\phi}_1\rangle$ block values during 
the simulation shown in Fig. \ref{fig:tp-a}. The figure shows the ``rotation'' of the 
field around the origin in proximity of the potential minima ring of radius 
$\bar{\Pi}_a\approx 0.977451$, for $N=8$ and 10, but not for $N=13$. Even for $N=8$ and 
10 the rotation was not symmetric (this would only be obtained in an extremely long 
simulation), which explains the not exactly zero value of the expectation value of 
field.}
\label{fig:Hbp}
\end{figure}

\begin{figure}[htbp]
\begin{center}
\includegraphics[width=12cm]{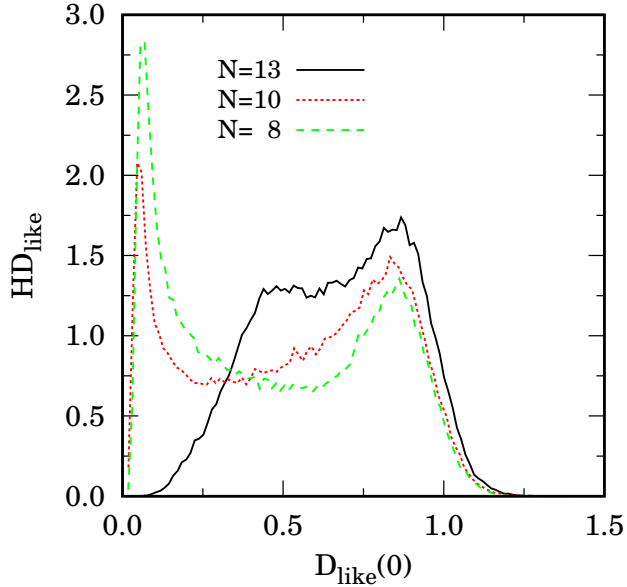}
\end{center}
\caption{(color online) Histogram of $D_{\rm like}(0)$ block values during the 
simulation shown in Fig. \ref{fig:tp-a}. The missing first peak in the $N=13$ data is 
due to the fact that the field didn't perform a full rotation around the origin as is 
shown by Fig. \ref{fig:Hbp}.}
\label{fig:HDl}
\end{figure}

\begin{figure}[htbp]
\begin{center}
\includegraphics[width=12cm]{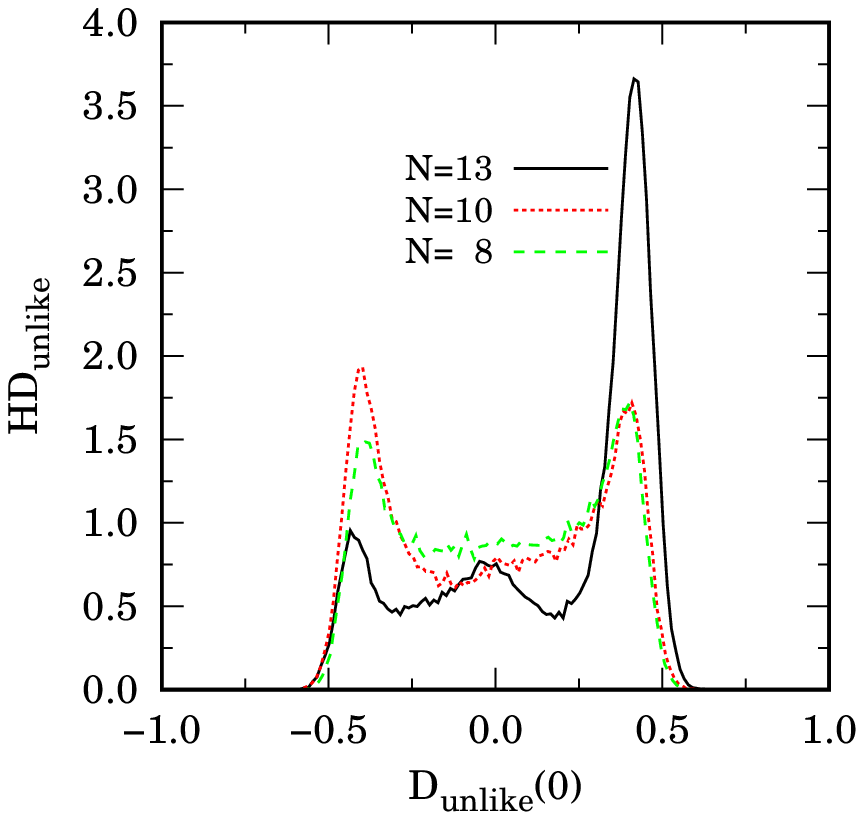}
\end{center}
\caption{(color online) Histogram of $D_{\rm unlike}(0)$ block values during the 
simulation shown in Fig. \ref{fig:tp-a}. The $N=13$ data presents a high asymmetry 
during the evolution of the simulation, which again signals that the simulation was not 
long enough.}
\label{fig:HDu}
\end{figure}

For the affine quantization case the circle of vacua has a radius $\bar{\Pi}_a$, which 
is now a function of $m$, $\bar{g}$, and $\bar{\Phi}$:
\bq \nonumber
\bar{\Pi}_a^2&=&\frac{4\bar{g}\bar{\Phi}^2-m^2}{12\bar{g}}+\\
&&\frac{(4\bar{g}\bar{\Phi}^2-m^2)^2}{12\bar{g}[162\bar{g}^2+\Xi+18\bar{g}^{2/3}(81\bar{g}^2+\Xi)^{1/3}]}+\frac{162\bar{g}^2+\Xi+18\bar{g}^{2/3}[81\bar{g}^2+\Xi]^{1/3}}{12\bar{g}},\\
\Xi&=&-m^6+12\bar{g}m^4\bar{\Phi}^2-48\bar{g}^2m^2\bar{\Phi}^4+64\bar{g}^3\bar{\Phi}^6,
\eq
where without loss of generality we assumed $\epsilon=0$. It is different from zero 
irrespectively from the values of the parameters, so symmetry is always broken.
In Fig. \ref{fig:tp-a} we show $D_{\rm like}(z)$ and $D_{\rm unlike}(z)$ as obtained 
for  $m=1, \bar{g}=1, \bar{\Phi}=1, L=3$ (so that $m/M=1$), $\epsilon=10^{-10}$ (the 
simulation results are not affected by $\epsilon$ as long as it is chosen sufficiently 
small), and three choices of $N$, in the affine scenario, for the $\bar{\phi}_i$ fields 
introduced in Eq. (\ref{eq:pb}). One can then see the approach to the continuum of the 
two-point functions of the affine model. Note, however, that now the region around 
$\bar{\phi}_i=0$ for $i=1,2$ is forbidden due to the affine $3/8$ diverging term in the 
potential density (see Eq. (\ref{eq:pb})), therefore the complex field in its 
``winding'' around the origin, in proximity of the potential minima ring, cannot take 
a ``shortcut'' through the ``mountain'' at the origin (the forbidden region) and this, 
in turn, is responsible for a loss of ergodicity and the appearance 
of systematic errors in addition to the usual statistical ones. It is then necessary an 
extremely long simulation (much longer than the average time for a ``round trip'' of 
the field), much longer than in the canonical case. Notice, moreover, that the action 
is penalized by the additional $a^{-s}$ factor which grows as we approach the continuum 
$a\to 0$. 
A possible solution would be to choose the field displacement $\delta$ larger than the 
diameter of the potential minima ring $2\bar{\Pi}_a$. But unfortunately this will 
not work because the kinetic energy term in the action doesn't allow the field to 
undergo big ``jumps''. In addition this would generate low acceptance ratios thereby 
slowing down the simulation. An alternative solution will be given in the next section.  
In our simulations, that were $M=10^7$ MC steps long, the expectation value of the 
field $\langle\bar{\phi}_1\rangle=\langle\bar{\phi}_2\rangle$ was equal to $-0.23(4)$ 
for $N=8$, to $-0.24(4)$ for $N=10$, and to $-0.784(9)$ for $N=13$. A non-zero value 
for the vacuum expectation of the field is due to the systematic errors described above 
and will eventually disappear in an extremely long simulation. 
From the figure we see how the two-point like function seems to be increasing with 
$N$, while the unlike one has a constant behavior fluctuating around the expected zero 
value. These results are still affected by the ergodicity systematic errors stemming 
from the ``winding'' random walk. In order to show this behavior, we calculated the 
histograms of the values for $\langle\bar{\phi}_1\rangle$ obtained by averaging over 
blocks of 100 MC steps during the simulation, that we call $H\bar{\phi}$, of 
$D_{\rm like}(0)$ that we call $HD_{\rm like}$, and of $D_{\rm unlike}(0)$, that we 
call $HD_{\rm unlike}$. 
The behavior of these histograms is shown in Figs. \ref{fig:Hbp}, \ref{fig:HDl}, and 
\ref{fig:HDu} respectively. From the histogram of Fig. \ref{fig:Hbp} we see how 
for $N=13$ the field did not have the chance of rotating around the origin and this 
explains the lack of the first peak in the histogram of Fig. \ref{fig:HDl}. 
We then conclude that the simulation for $N=13$ was not long enough. And this is 
responsible for the high value of the two-point like function observed for $N=13$, as 
shown in Fig. \ref{fig:tp-a}. In order to obtain a fully symmetric rotation of the 
field random walk around the origin we would clearly need an extremely long 
simulation. Nonetheless from the partial results of our long simulation we 
can gather a flavor of the convergence of the two-point functions in the affine case in 
the continuum limit at finite volume.


These results, albeit partial in their nature, give to affine quantization a role 
as a method producing meaningful quantum field theories even when, as we have already 
seen in our previous works \cite{Fantoni2020,Fantoni2020a,Fantoni2020b}, the more 
common canonical quantization fails. Moreover with the scaling used in Eq. 
(\ref{eq:pb}) the field theory doesn't suffer from the unpleasant feature of a 
diverging vacuum expectation value of the field in the continuum limit, which was 
observed in Ref. \cite{Fantoni2020b}. 

\section{Exponential representation of the complex field} 
\label{sec:exponential}

In order to solve the ergodicity breakdown problem encountered in the previous section 
for the affine case we decided to rewrite our path integral in terms of the fields 
$\rho(x)$ and $\theta(x)$ such that $\bar{\varphi}(x)=\rho(x)\exp[i\theta(x)]$. Eq. 
(\ref{eq:pb}) may be rewritten as follows
\bq \nonumber
\frac{S^{(a)}[\rho,\theta]}{a^{-s}a^n}\approx\frac{1}{2a^2}&\sum_{x,\mu}&\left\{\left[\rho(x)-\rho(x+e_\mu)\right]^2+\rho^2(x)\left[\theta(x)-\theta(x+e_\mu)\right]^2\right\}+\\ \label{eq:pc}
&\sum_x&\left\{\frac{3}{8}\frac{1}{\rho^2(x)+\bar{\epsilon}}+ 
\frac{1}{2}m^2\rho^2(x)+\bar{g}\left[\rho^2(x)-\bar{\Phi}^2\right]^2\right\},
\eq
and the path integral over $\rho\in[0,\infty]$ and $\theta\in[-\infty,\infty]$ will not 
suffer anymore from the ergodicity problem. In the Metropolis algorithm we will now 
have acceptance when $\exp[-(S'-S)]\prod_x\rho'(x)/\rho(x)>\eta$ where the primed 
quantities are the newly generated ones and as usual $\eta$ is a pseudo random number 
in $[0,1]$. The modulus displacement move, 
$\rho\rightarrow\rho'=\rho+(2\eta-1)\delta_\rho$, is rejected whenever $\rho'<0$. And 
the argument displacement move is chosen purposely asymmetric, 
$\theta\rightarrow\theta'=\theta+\eta\delta_\theta$, in order to allow for the required 
rotation and break the symmetry. This transition rule for the argument will not violate 
the detailed balance, required by the Metropolis algorithm, as long as the maximum 
displacement is chosen $\delta_\theta\geq 2\pi$ so that the probability to go from an 
angle $\theta_A$ to $\theta_B$ will be equal to the one to return to $\theta_A$ from 
$\theta_B$ always using counterclockwise rotations. 

\begin{figure}[htbp]
\begin{center}
\includegraphics[width=12cm]{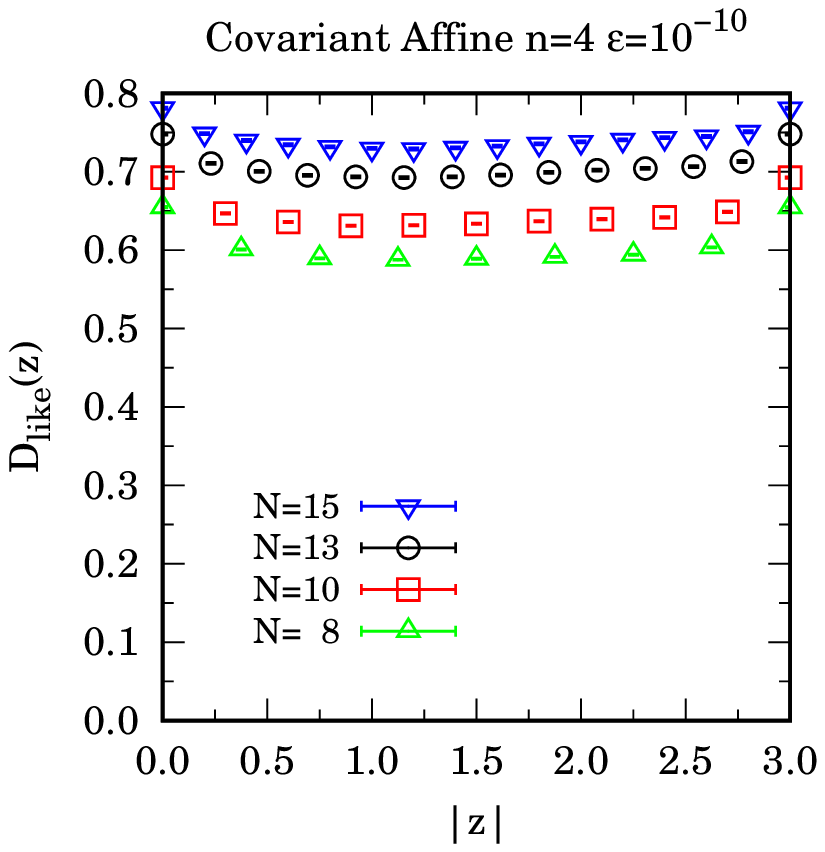}\\
\includegraphics[width=12cm]{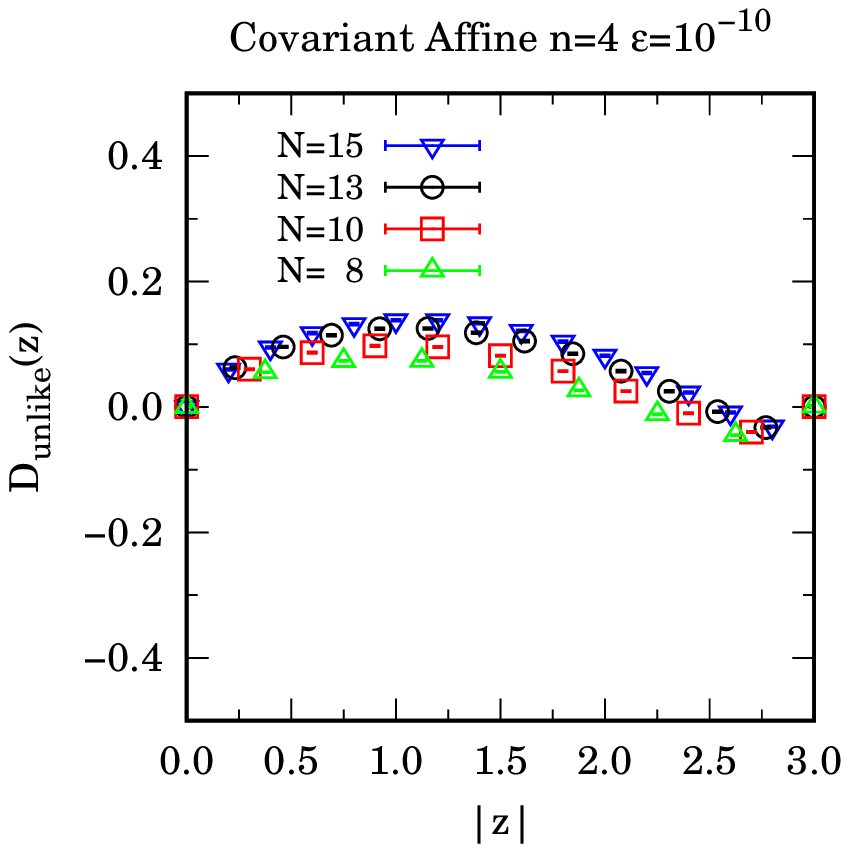}
\end{center}
\caption{(color online) Two-point functions, $D_{\rm like}(z)$ (top panel) and 
$D_{\rm unlike}(z)$ (bottom panel), of Eq. (\ref{eq:tp}) for the complex scalar Higgs 
field $\bar{\varphi}=\bar{\phi}_1+i\bar{\phi}_2=\rho\exp(i\theta)$ subject to affine  
quantization with a self-interaction potential density of the form 
$V=\frac{1}{2}m^2\rho^2+g(\rho^2-\Phi^2)^2$ in Eq. (\ref{eq:a-action}) with 
$m=1, \bar{g}=1, \bar{\Phi}=1, L=3, \epsilon=10^{-10}$ in Eq. (\ref{eq:pc}) and  
increasing $N=8,10,13,15$. The simulation used $M=10^6$ MC steps. On the abscissa axis 
we have $|z|=\sqrt{z_0^2+z_1^2+\ldots+z_s^2}$ which is a length.}
\label{fig:tp-exp}
\end{figure}

In Fig. \ref{fig:tp-exp} we show $D_{\rm like}(z)$ and $D_{\rm unlike}(z)$ as obtained 
for  $m=1, \bar{g}=1, \bar{\Phi}=1, L=3,\epsilon=10^{-10}$, and four choices of 
increasing $N$, in the affine scenario, for the fields $\bar{\phi}_1=\rho\cos\theta$ 
and $\bar{\phi}_2=\rho\sin\theta$. The simulations, an order of magnitude shorter than 
the one of Fig. \ref{fig:tp-a}, rapidly converged and we had vanishing 
$\langle\bar{\phi_i}\rangle$ as required. From the figure we can see how the symmetry 
$z\to L-z$ appears to be broken in both two-point functions. In particular the unlike 
one appears to be oscillating close to the value of zero. This can be seen as an 
artifact due to the chosen asymmetric expression for the kinetic part of the primitive 
approximation. The two-point functions, that are now well converged, seem to have a 
well defined continuum limit $N\to\infty$. In fact the difference between 
$D_{\rm like}(|z|=L/2)$ from $N=10$ and $N=8$ is 0.043 but the one from $N=15$ and 
$N=13$ is 0.036. This supports the conclusion that affine quantization leads to a well 
defined field theory. This is also supported by looking at the renormalized mass and 
coupling constant \cite{Fantoni2020a}: 
$m_R=0.101748(8), \bar{g}_R=1.50000(1)$ for $N=8$, 
$m_R=0.097307(8), \bar{g}_R=1.50000(2)$ for $N=10$, 
$m_R=0.08949(4),  \bar{g}_R=1.50000(3)$ for $N=13$,
$m_R=0.08398(6),  \bar{g}_R=1.49997(4)$ for $N=15$. We can see how the renormalized 
coupling constant remains constant upon the increase of $N$.

%

\section{Conclusions} 
\label{sec:conclusions}

Summarizing, in this work we studied, through Monte Carlo simulations, the two-point 
function of a classical Euclidean covariant complex scalar field of mass $m$ subject to 
the Higgs Mexican-hat potential in four space-time dimensions, treated either with 
canonical quantization and with affine quantization. And we analyzed the continuum 
limit at finite fixed volume. The finite volume constraint rules out the formation of 
the massless Goldstone boson due to the spontaneous symmetry breaking of the continuous 
phase symmetry $\varphi(x)\to e^{i\theta(x)}\varphi(x)$ that we continue to observe in 
the simulations even if only as a smooth transition (free energies in finite volume 
systems are always analytic).

We first studied the path integral in the two real fields $\phi_1$ and $\phi_2$ with 
$\varphi=\phi_1+i\phi_2$ through standard Metropolis \cite{Metropolis} simulations.
In the canonical case we found rapidly converging simulations: the unlike two-point 
function is zero everywhere and the like one shows the approach to the continuum 
through a diverging value at the origin. It is periodic of periodicity $L$ and 
satisfies the symmetry $z\to L-z$ as it should. It has a minimum at half simulation box 
$|z|=L/2$ close to zero, indicating that the scalar field theory is in the unbroken 
phase under canonical quantization, at the chosen couplings and dimension.

In the affine case we found that due to the appearance of the forbidden region around 
the origin $\varphi\approx 0$, the ergodicity of the random walk is broken. Once the 
field spontaneously breaks the symmetry falling in the circle of vacua, it can only 
rotate around the peak in the potential at the origin. Therefore very long simulations 
are necessary in order to find reliable results for the expectation values. More so 
approaching the continuum. This suggested to change variables from $\phi_1$ and 
$\phi_2$ to the modulus $\rho$ and the argument $\theta$ of the complex field, with 
$\varphi=\rho\exp(i\theta)$ and choose an asymmetric transition rule for the argument 
move in the Metropolis algorithm in order to allow only for counterclockwise rotations 
around the origin. This proved an effective way to overcome the ergodicity problem 
encountered previously, and the simulations converged quickly. 

The approach to the continuum appears to be well behaved also for the affine case where 
the unlike two-point function continues to be everywhere close to zero and the like 
one develops a minimum at half simulation box higher than the one observed in the 
canonical case indicating that the system under affine quantization is in the broken
phase. Therefore we can say that affine quantization produces a meaningful quantum 
field theory. It would be interesting to carry on a detailed and systematic study of 
the approach to the continuum of the renormalized coupling constant in order to 
understand whether the affine approach is able to produce a non-trivial 
\cite{Wilson1974,Frohlich1982,Weisz1987,Weisz1988,Weisz1989} interacting field theory 
in the continuum limit also for the present case of a scalar complex field subject to 
the Higgs potential, as was done in our previous works for scalar real fields 
\cite{Fantoni2020,Fantoni2020a}. This would solve the problem of the believed 
triviality of the canonical Higgs particle in four space-time dimensions.

\appendix


%

\end{document}